\documentclass[12pt,preprint]{aastex}

\newcommand{\IRIS}{{\em IRIS}}
\newcommand{\SDO}{{\em SDO}}
\newcommand{\GOES}{{\em GOES}}
\newcommand{\SOHO}{{\em SOHO}}

\shorttitle{Spectroscopic Observations of an X-shaped Flare}
\shortauthors{Li et al.}

\begin{document}

\title{Spectroscopic Observations of Magnetic Reconnection and \\Chromospheric Evaporation in an X-shaped Solar Flare}

\author{Y. Li$^{1,2,3,4,5}$, M. Kelly$^{3}$, M. D. Ding$^{2,5}$, J. Qiu$^{3}$, X. S. Zhu$^{4}$, W. Q. Gan$^{1}$}
\affil{$^1$Key Laboratory of Dark Matter and Space Astronomy, Purple Mountain Observatory, Chinese Academy of Sciences, Nanjing 210008, China}
\affil{$^2$School of Astronomy and Space Science, Nanjing University, Nanjing 210023, China}
\affil{$^3$Department of Physics, Montana State University, Bozeman, MT 59717, USA}
\affil{$^4$CAS Key Laboratory of Solar Activity, National Astronomical Observatories, Beijing 100012, China}
\affil{$^5$Key Laboratory for Modern Astronomy and Astrophysics (Nanjing University), Ministry of Education, Nanjing 210023, China}

\begin{abstract}
We present observations of distinct UV spectral properties at different locations during an atypical X-shaped flare (SOL2014-11-09T15:32) observed by the {\em Interface Region Imaging Spectrograph} (\IRIS). In this flare, four chromospheric ribbons appear and converge at an X-point where a separator is anchored. Above the X-point, two sets of non-coplanar coronal loops approach laterally and reconnect at the separator. The \IRIS~slit was located close to the X-point, cutting across some of the flare ribbons and loops. Near the location of the separator, the Si {\sc iv} 1402.77 \AA~line exhibits significantly broadened line wings extending to 200 km s$^{-1}$ but an unshifted line core. These spectral features suggest the presence of bidirectional flows possibly related to the separator reconnection. While at the flare ribbons, the hot Fe {\sc xxi} 1354.08 \AA~line shows blueshifts and the cool Si {\sc iv} 1402.77 \AA, C {\sc ii} 1335.71 \AA, and Mg {\sc ii} 2803.52 \AA~lines show evident redshifts up to a velocity of 80 km s$^{-1}$, which are consistent with the scenario of chromospheric evaporation/condensation.
\end{abstract}


\noindent

 \section{Introduction}

Solar flares (see a recent review by \citealt{flet11}) are energetic events in the solar atmosphere, which are believed to be powered by magnetic reconnection in the corona \citep{prie02,shib11}. The energy released by reconnection usually heats the local plasma and accelerates particles. Through thermal conduction and/or non-thermal particle beams, the energy is then transported downward to the lower atmosphere. Consequently, the chromospheric plasma is heated and emits an enhanced radiation, which outlines the flare ribbons. An impulsive energy deposition leads to a local pressure excess that drives the heated plasma up into the corona, referred to as chromospheric evaporation \citep{neup68,hira74,acto82}. The evaporated hot plasma fills the flare loops which are clearly visible in soft X-ray and EUV passbands. 

Magnetic reconnection, as the dominant energy release mechanism in flares, has been reported in spectroscopic observations from different instruments. Using the Solar Ultraviolet Measurements of Emitted Radiation (SUMER; \citealt{wilh95}) spectrometer on the {\em Solar and Heliospheric Observatory} (\SOHO), \cite{inne03a,inne03b} observed evident blue-wing enhancements at 800--1000 km s$^{-1}$ in the Fe {\sc xxi} 1354.08 \AA~line on the top of flare arcades (viewed at the solar limb), which are associated with supra-arcade (or reconnection) downflows \citep{mcke99}. A high blueshifted jet (with a velocity up to 600 km s$^{-1}$ along the line of sight) and a redshifted jet ($\sim$300 km s$^{-1}$) were also recorded by SUMER in the Fe {\sc xix} 1118.07 \AA~line near the top of erupting loops, both of which were explained as reconnection outflows \citep{wang07}. In the era of the {\em Hinode} EUV Imaging Spectrometer (EIS; \citealt{culh07}), \cite{hara11} reported reconnection outflows with a velocity of $\sim$200--400 km s$^{-1}$ in the Fe {\sc xxiv} 192.03 \AA~and Ca {\sc xvii} 192.86 \AA~lines as well as reconnection inflows with a velocity of $\sim$20 km s$^{-1}$ in the Fe {\sc xii} 195.12 \AA~and Fe {\sc x} 184.54 \AA~lines around the loop-top region. In addition, \cite{simo15} detected high redshifts (40--250 km s$^{-1}$) in the EIS Fe {\sc xxiv} 192.03 \AA~and Fe {\sc xii} 192.39 \AA~lines at a coronal source in terms of reconnection downflows. Recently, using the high-resolution UV spectra from the {\em Interface Region Imaging Spectrograph} (\IRIS; \citealt{depo14}), \cite{tian14} reported a large redshift ($\sim$125 km s$^{-1}$) in the Fe {\sc xxi} 1354.08 \AA~line on the cusp-shaped structure and interpreted the redshift as a signature of reconnection downflows. The authors also observed a downward-moving blob as manifested by a greatly enhanced redshifted component at $\sim$60 km s$^{-1}$ in the Si {\sc iv} 1402.77 \AA~line. Moreover, \cite{reev15} found intermittent fast flows of 200 km s$^{-1}$ in dome-shaped coronal loops in the \IRIS~Si {\sc iv} 1393.76 \AA~line and considered the fast flows as a result of magnetic reconnection between an erupting prominence and the pre-existing overlying magnetic field. Reconnection signatures were also observed in the cooler H$\alpha$ and Ca {\sc ii} 8542 \AA~lines by the ground-based Fast Imaging Solar Spectrograph (FISS; \citealt{chae13}) manifested as bidirectional outflows with velocities of $\pm$(70--80) km s$^{-1}$ \citep{hong16}. Note that evidence of magnetic reconnection has also been found in small-scale explosive events \citep{dere89}. For example, broad non-Gaussian Si {\sc iv} line profiles with both wings extending to hundreds of km s$^{-1}$ have been observed and explained as the consequence of bidirectional reconnection jets \citep{dere91,inne97,inne15,tian16}.

Chromospheric evaporation, i.e., a dynamic response to the energy deposition from magnetic reconnection, can be detected by Doppler shift measurements in spectral lines. In general, the evaporated (or upward) plasma motions generate blueshifts (or blueshifted components) in soft X-ray and EUV lines. Based on momentum balance \citep{canf87,canf90}, chromospheric evaporation is usually accompanied by a compression of chromospheric plasma, called chromospheric condensation, which produces redshifts (or red-wing enhancements)  in some relatively cool lines. Blueshifts (redshifts) caused by chromospheric evaporation (condensation) have been reported in a large number of studies, for example, blueshifted components with velocities of 200--400 km s$^{-1}$ in the Ca {\sc xix} 3.18 \AA~line \citep{anto82,anto85,anto83,zarr88,wuls94,ding96} from the Bent and Bragg Crystal Spectrometer (BCS; \citealt{acto80}) on board the {\em Solar Maximum Mission} ({\em SMM}) and {\em Yohkoh}/BCS (\citealt{culh91}), blueshifts of 60--300 km s$^{-1}$ in the Fe {\sc xix} 592.23 \AA~line \citep{teri03,teri06,delz06,bros04} from the Coronal Diagnostic Spectrometer (CDS; \citealt{harr95}) on board \SOHO, and redshifts of tens of km s$^{-1}$ in some chromospheric and transition region lines (like H$\alpha$, He {\sc ii} 303.78 \AA, O {\sc iii} 599.59 \AA, and O {\sc v} 629.73 \AA; \citealt{wuls94,ding95,czay99,teri03,teri06,bros03,kami05,delz06}). In particular, blueshifts and redshifts can appear at a given flaring location in different emission lines, as observed by {\em Hinode}/EIS \citep{mill09,chen10,ying11,dosc13}, confirming the coexistence of chromospheric evaporation and condensation. Recently, \IRIS~also observed blueshifts of hundreds of km s$^{-1}$ in the hot Fe {\sc xxi} 1354.08 \AA~line and redshifts of tens of km s$^{-1}$ in the cool Si {\sc iv} 1402.77 \AA~(or 1393.76 \AA), C {\sc ii} 1335.71 \AA~(or 1334.53 \AA), and Mg {\sc ii} 2803.52 \AA~(or 2796.35 \AA~and 2791.59 \AA)~lines at flare ribbons or kernels, which have been explained by chromospheric evaporation and condensation, respectively \citep{tian14,tian15,youn15,ying15,grah15,bros15,batt15,lido15,poli15,poli16,sady15,sady16,dudi16}. Note that the Fe {\sc xxi} line is usually blueshifted as a whole, while the Mg {\sc ii}, C {\sc ii}, and Si {\sc iv} lines typically only show a red-wing enhancement. According to the observed Doppler shifts (or depending on the heating rate), chromospheric evaporation can be divided into two types: gentle evaporation and explosive evaporation \citep{fish85a,fish85b,fish85c,mill06a,mill06b}. When the hot lines (such as Fe {\sc xxi}, Fe {\sc xix}, and Ca {\sc xix}) show blueshifts and the cool lines (such as H$\alpha$, He {\sc ii}, O {\sc iii}, and Si {\sc iv}) show redshifts, this case is referred to as explosive evaporation. When only blueshifts are detected, this corresponds to gentle evaporation. Both types of evaporation have been observed in a single flare \citep{bros09,ying11}. In addition, explosive evaporation could occur in major flares \citep{mill06b,vero10} as well as in microflares \citep{bros10,chen10}.

In this paper, we present spatio-temporal variation of the UV spectra of Si {\sc iv}, C {\sc ii}, and Mg {\sc ii} observed by \IRIS~for an atypical X-shaped flare, in which magnetic reconnection takes place at a separator \citep{ying16}. The separator reconnection creates four chromospheric ribbons that converge at an X-point as revealed by the \IRIS~slit-jaw 1330 \AA~images (SJIs). Accordingly, two sets of non-coplanar flare loops take part in the reconnection, as shown in the EUV images from the Atmospheric Imaging Assembly (AIA; \citealt{leme12}) on board the {\em Solar Dynamics Observatory} (\SDO). The \IRIS~slit was located near the X-point, capturing some of the X-shaped flare ribbons and also non-coplanar flare loops involved in the separator reconnection. From the observed spectra, we detect convincing upward and downward reconnection outflows near the location of the separator in the Si {\sc iv} line, which are rarely reported in previous flare studies. In addition, we find some interesting features corresponding to chromospheric condensation at flare ribbons, such as entirely redshifted Si {\sc iv} line profiles.

\section{Observations and Data Reduction}
\label{sec-obs}

\begin{figure*}
\centering
\includegraphics[width=13.5cm]{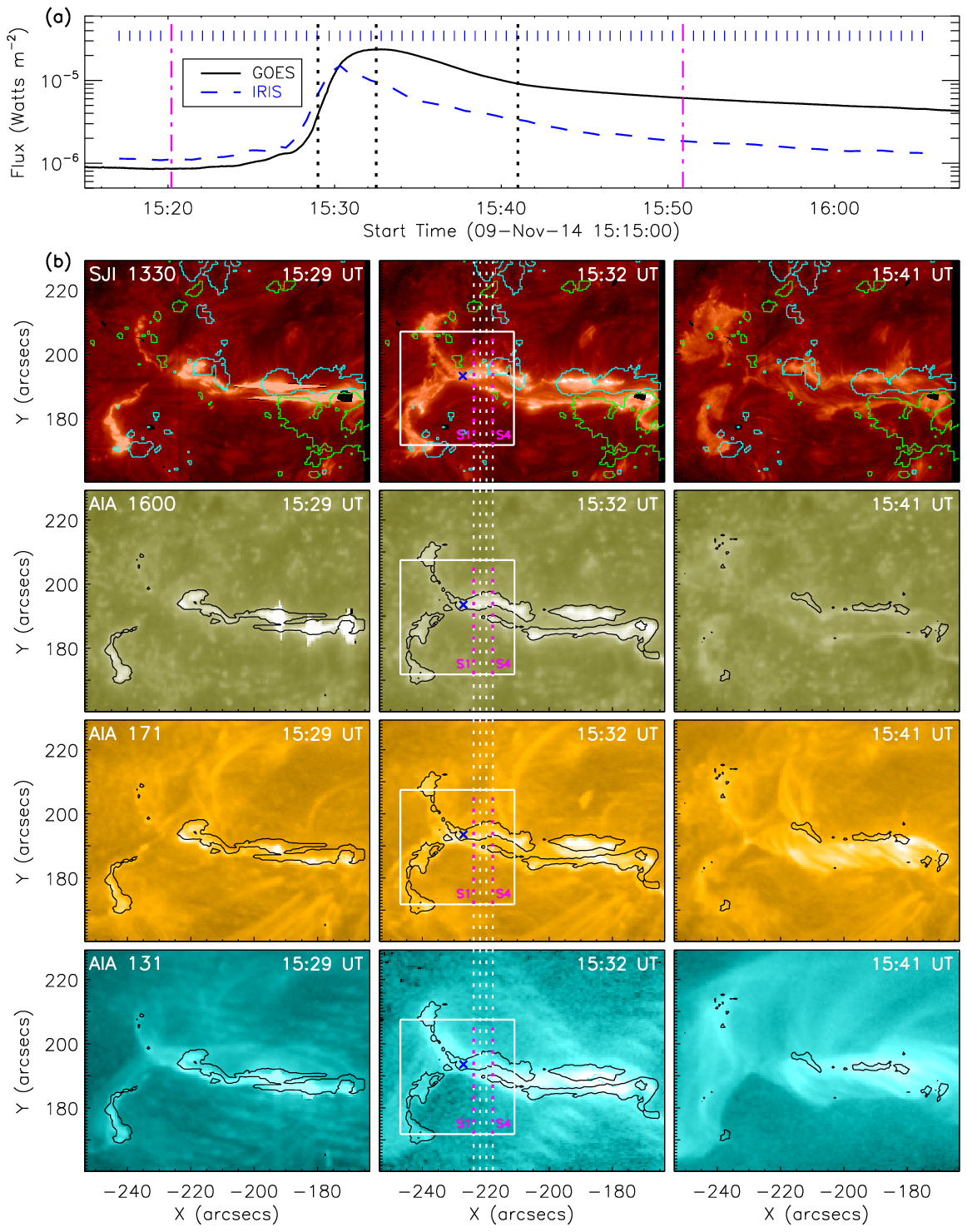}
\caption{{\small Overview of the X-shaped flare event. (a) Light curves of \GOES~1--8 \AA~for the whole Sun and of SJI 1330 \AA~(arbitrary scale) integrated over the region as shown in panel (b). The short vertical blue lines mark the observing times of SJIs at 1330 \AA. The two magenta dash-dotted lines indicate the time range as shown in Figures \ref{fig-siw0}--\ref{fig-fel}. The three black dotted lines mark the times of each image in panel (b) with the middle line denoting the flare peak. (b) SJIs at 1330 \AA~and AIA 1600, 171, and 131 \AA~images during the flare. The four vertical dotted lines indicate the scan steps of the \IRIS~slit, of which the first and fourth ones (S1 and S4) are studied in detail. The white box denotes the region around the X-point (marked by the blue cross), as shown in Figures \ref{fig-fpt}--\ref{fig-siw0} and \ref{fig-siw3}. Superimposed on SJIs at 1330 \AA~are the contours of  magnetic polarities at $+$500 G (cyan) and $-$500 G (green). Also plotted on the AIA 1600, 171, and 131 \AA~images are the contours of the SJI 1330 \AA~intensity of 25 times the average quiescent-Sun level, denoting the chromospheric ribbons.}}
\label{fig-obs}
\end{figure*}

The X-shaped flare is a \GOES~M2.3 event that occurred on 2014 November 9 in the active region NOAA 12205 near the disk center (N14E11). It started at 15:24 UT, peaked at 15:32 UT, and lasted until 16:05 UT (see the \GOES~1--8 \AA~soft X-ray flux in Figure \ref{fig-obs}(a)). \IRIS~observed this flare from 15:17 UT to 16:05 UT (indicated by the short vertical blue lines in Figure \ref{fig-obs}(a)), covering the entire rise phase and almost all the decay phase of the flare. The \IRIS~SJIs at 1330 \AA~show that four chromospheric ribbons appear in a quadrupolar magnetic field and converge at an X-point (Figure \ref{fig-obs}(b) and Animation 1) where a separator is anchored \citep{ying16}. In addition, two sets of non-coplanar coronal loops approach laterally and reconnect at the separator, as revealed by the AIA images (Figure \ref{fig-obs}(b) and Animation 1). \IRIS~also observed spectra over a small region (marked by the white dotted lines in Figure \ref{fig-obs}(b)) to the west of the X-point with an offset of $\sim$4\arcsec. This small region contains some of the flare ribbons as well as coronal loops that are associated with the separator reconnection \citep{ying16}.

\IRIS~provides high-resolution SJIs as well as spectra using a slit in the near-ultraviolet (NUV, 2783--2834 \AA) and far-ultraviolet (FUV, 1332--1358 \AA~and 1389--1407 \AA) wavelengths. The slit has a width of $0.\!\!^{\prime\prime}33$ and is located in the middle of the SJIs that have a pixel scale of $0.\!\!^{\prime\prime}167$. For this flare, \IRIS~observed an area of 119\arcsec$\times$119\arcsec~in SJIs at 1330, 2796, and 2832 \AA~with a cadence of 37 s, the former two of which are sensitive to the plasma of upper chromosphere and the latter one to the upper photosphere \citep{depo14}. The slit scanned a small area of 6\arcsec$\times$119\arcsec~with four steps (i.e., each step moves 2\arcsec~across the slit). It took 37 s in each run with an exposure time of 8 s at each step. In the present study, we focus on SJIs at 1330 \AA~for the X-point region (with an area of 36\arcsec$\times$36\arcsec; see the white box in Figure \ref{fig-obs}(b)) as well as the spectra from the first and fourth steps (referred to as S1 and S4 hereafter and marked by the two magenta dotted lines in Figure \ref{fig-obs}(b)) within the time range of 15:20--15:50 UT (denoted by the two magenta dash-dotted lines in Figure \ref{fig-obs}(a)). Note that S1 is the closest while S4 is the farthest to the X-point. We use the \IRIS~level 2 data that have been processed with the subtraction of dark current as well as the corrections for flat field, geometry, and wavelength.

The spectra studied here include the Mg {\sc ii} line at 2803.52 \AA~(with a formation temperature of $\sim$10$^{4.0}$ K), the C {\sc ii} line at 1335.71 \AA~($\sim$10$^{4.3}$ K), the Si {\sc iv} line at 1402.77 \AA~($\sim$10$^{4.8}$ K), and the Fe {\sc xxi} line at 1354.08 \AA~($\sim$10$^{7.0}$ K). The chromospheric Mg {\sc ii} and C {\sc ii} lines are optically thick (i.e., formed through a complex radiative transfer process) and usually show a central reversal in the line core. We therefore adopt a moment method to analyze these two lines and obtain the spectral parameters, i.e., total intensity (the zeroth order moment), line shift (the first order moment), and line width (the second order moment). The transition region Si {\sc iv} line is usually regarded as an optically thin line that can be fitted by a Gaussian function \citep{bran15}. It should be noted that, in some observations, the Si {\sc iv} line profiles significantly deviate from a Gaussian shape. Considering these, we first do the moment analysis on all of the observed Si {\sc iv} profiles; then, in some specific places that exhibit line profiles with Gaussian shapes, we also implement a single or multiple Gaussian fitting. For the coronal Fe {\sc xxi} line that is optically thin, we just apply a Gaussian fitting to derive the spectral parameters. Note that the Fe {\sc xxi} line is blended with some other weak lines, thus we adopt a multiple Gaussian fitting to separate the Fe {\sc xxi} component from the other components \citep{ying15}. To calculate the Doppler velocity from the line shift, we first determine the reference line center by averaging the observed line centers before the flare onset for the Mg {\sc ii}, C {\sc ii}, and Si {\sc iv} lines\footnote{Assuming no shifts in the cold lines (e.g., S {\sc i}, Fe {\sc i}, and Mn {\sc i}) before the flare onset should be the best way to determine the rest wavelength. However, no cold lines are clearly visible in the C {\sc ii} and Si {\sc iv} spectral windows for the data used here. Note that we do see some cold lines in the Mg {\sc ii} spectral window. Here we compare the Doppler velocities of Mg {\sc ii} derived from two methods, i.e., via the cold Mn {\sc i} 2801.91 \AA~line and via the average line centers, and only find a small difference of 0.56 km s$^{-1}$ between them.}. For the hot Fe {\sc xxi} line that cannot be seen before the flare onset, we use the theoretical line center, i.e., 1354.08 \AA. This value is very close to the reference line centers independently determined by \cite{youn15} and \cite{bros15}. The uncertainty in the Doppler velocity for all the lines is estimated to be less than 10 km s$^{-1}$ \citep{ying15}.

In this study, we also use the AIA EUV and UV images with a pixel scale of $0.\!\!^{\prime\prime}6$ and cadences of 12 s and 24 s, respectively. The images of AIA and \IRIS~are co-aligned by comparing the sunspot features visible in the AIA 1700 \AA~and SJI 2832 \AA~images (as shown in Animation 2). The \IRIS~SJIs themselves are also co-aligned by correcting a drift of $\sim$2\arcsec~in the X-direction throughout the flare (see the slit position in Animation 2). The uncertainty in the co-alignments of different images is estimated to be $\sim$1\arcsec.

\section{Spatial Context of the Event}
\label{sec-spa}

\begin{figure*}
\centering
\includegraphics[width=15cm]{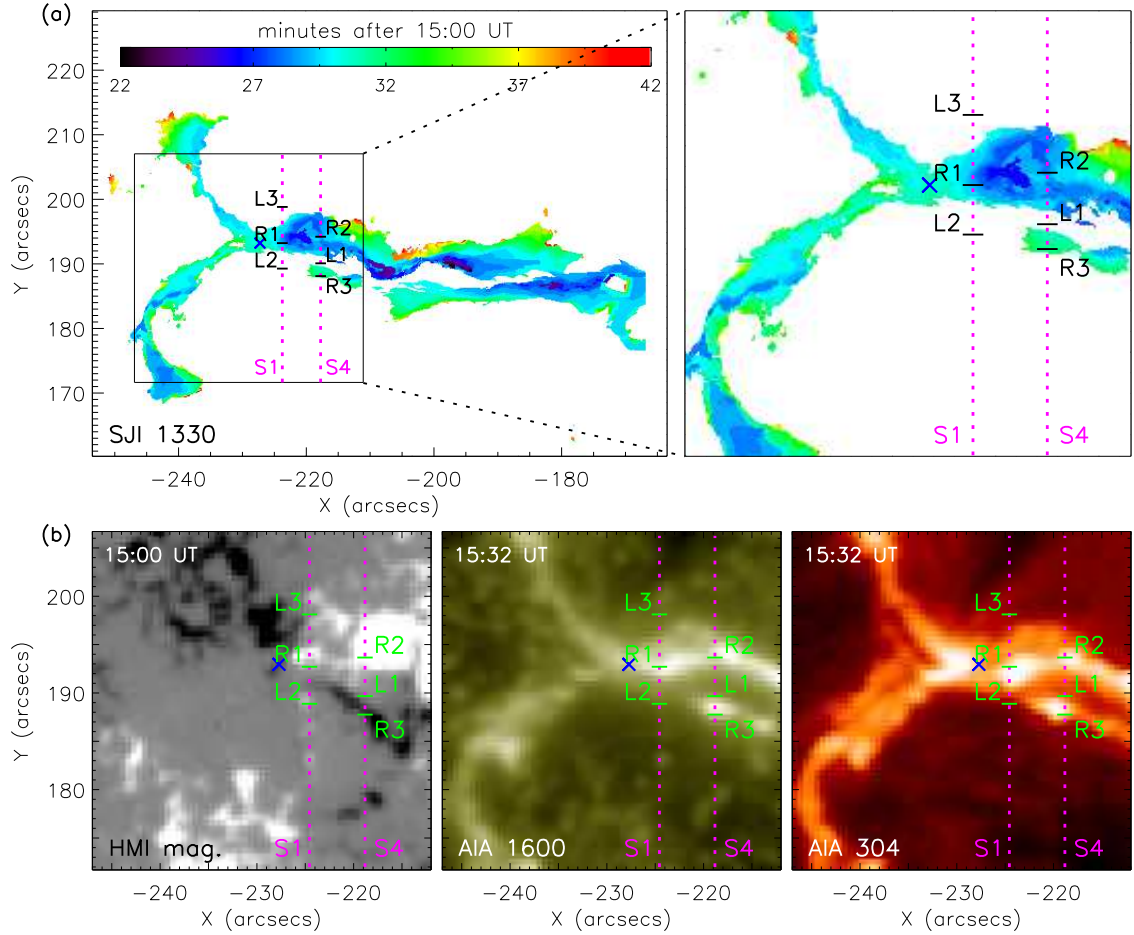}
\caption{{\small (a) Evolution of the flare ribbon brightenings observed in SJI 1330 \AA~(left) and a zoomed-in view of the X-point region (right). The first and fourth steps (S1 and S4) of the \IRIS~slit are plotted in magenta lines on which some sample pixels are selected for study (R1--R3 are ribbon locations and L1--L3 are locations outside the ribbons). (b) HMI magnetogram before the flare and AIA 1600 and 304 \AA~images around the flare peak time for the X-point region. The X-point is indicated by a blue cross.}}
\label{fig-fpt}
\end{figure*}

\begin{figure*}
\centering
\includegraphics[width=16cm]{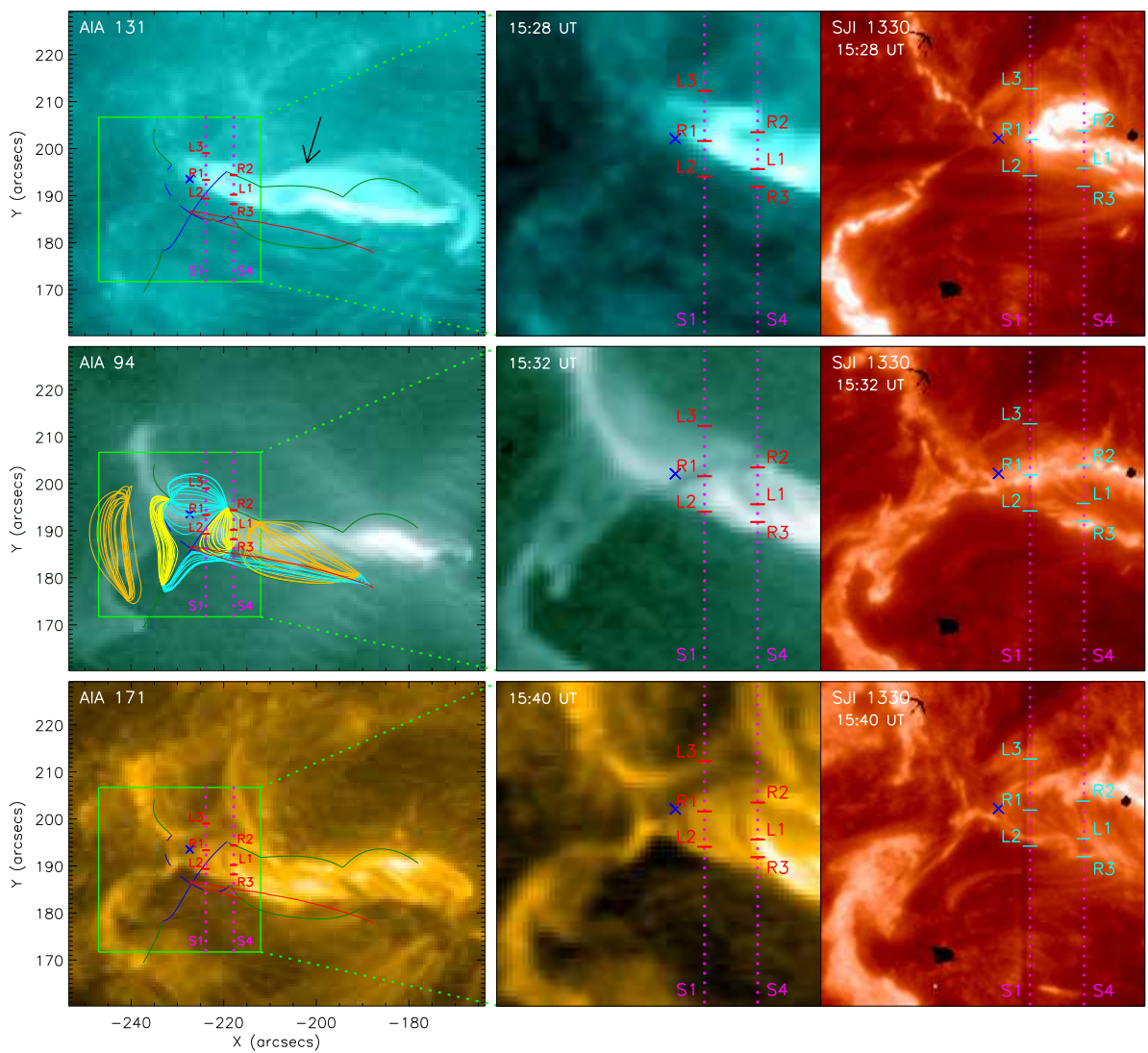}
\caption{{\small AIA 131, 94, and 171 \AA~images for the whole flaring region (left), zoomed-in view of the X-point region (middle), and SJIs at 1330 \AA~(right) in different flare phases. The slit steps and sample locations are marked in each panel (the same as in Figure \ref{fig-fpt}). The blue cross indicates the X-point. In the left column, the blue and dark green lines show the modeled separatrix traces, and the red line indicates the separator (i.e., intersection of separatrices) connecting to null points. The cyan, yellow, and orange lines in the AIA 94 \AA~image highlight some representative field lines (cyan for the pre-reconnection domain; yellow and orange for the post-reconnection domain). The black arrow in the AIA 131 \AA~image marks the erupted flux rope in the event.}}
\label{fig-lps}
\end{figure*}

The morphological evolution of this X-shaped flare has been well described in \cite{ying16}. In short, the four chromospheric ribbons approach each other and converge at the X-point around the flare peak time; then they move outward with the two ribbons on the right separating away from the polarity inversion line (the ribbon motion pattern\footnote{For the ribbon brightenings in SJI 1330 \AA, we select the flaring pixels whose intensity is enhanced to be more than 25 times the intensity of the quiescent Sun for more than two minutes (see \citealt{ying16} for more details).} is also shown in Figure \ref{fig-fpt}(a)). The observed ribbon motions as well as the reconstructed magnetic topology\footnote{The method of Magnetic Charge Topology \citep{long05} is used to produce the topological model, which has an uncertainty of $\sim$15\% in the connectivity of extrapolated field lines.} (see Figure \ref{fig-lps} and more details in \citealt{ying16}) suggest that magnetic reconnection takes place at a separator connecting to the X-point. More specifically, the inward and outward motions of ribbon brightenings illustrate that the reconnection occurs along a curved separator (or current sheet) which consists of a vertical part above the X-point and a horizontal part extending to the right (see the sketched Figure 4 in \citealt{ying16}). The \IRIS~slit cut across parts of the flare ribbons near the X-point (see Figure \ref{fig-fpt} and Animation 1). It also crossed some flaring loops and perhaps the curved separator as well (Figure \ref{fig-lps} and Animation 1). These provide us an opportunity to study the dynamics near the location of separator reconnection and at the flare ribbons from the observed UV spectra.

From Figure \ref{fig-fpt} and Animation 1, it is seen that the first step of the slit, i.e., S1, is 6\arcsec~closer to the X-point than the fourth step, S4, and that S1 cuts across only one of the flare ribbons while S4 cuts both ribbons to the right of the X-point. To investigate the ribbon dynamics, we present space-time diagrams of different spectral parameters for S1 and S4 (Figures \ref{fig-siw0}--\ref{fig-cmg3}) as well as some typical line profiles (Figures \ref{fig-pfr} and \ref{fig-fer}) at three ribbon pixels R1, R2, and R3 (R1 is located at the north ribbon cut by S1, and R2 and R3 are located at the north and south ribbons cut by S4, respectively). In addition, we show the line profiles (Figures \ref{fig-pfl} and \ref{fig-fel}) for some other pixels outside the flare ribbons, labeled as L1, L2, and L3 (L1 is on S4 and L2 and L3 are on S1), which display quite different dynamic features. The three pixels L1--L3 are presumably located on some loop structures or even at the separator as suggested in Figure \ref{fig-lps}. It should be noted that there is an overlapping effect along the line of sight and some of the sample pixels may correspond to different structures in different passbands. In the following section, we first present the spectral features at the flare ribbons, particularly at locations R1--R3 (S\ref{sec-eva}); then we show the distinct dynamic features at locations L1--L3 outside the flare ribbons (S\ref{sec-rec}).

\section{Results}

\subsection{Spectral Features at the Flare Ribbons}
\label{sec-eva}

\subsubsection{General Picture from the Moment Analysis}

\begin{figure*}
\centering
\includegraphics[width=16cm]{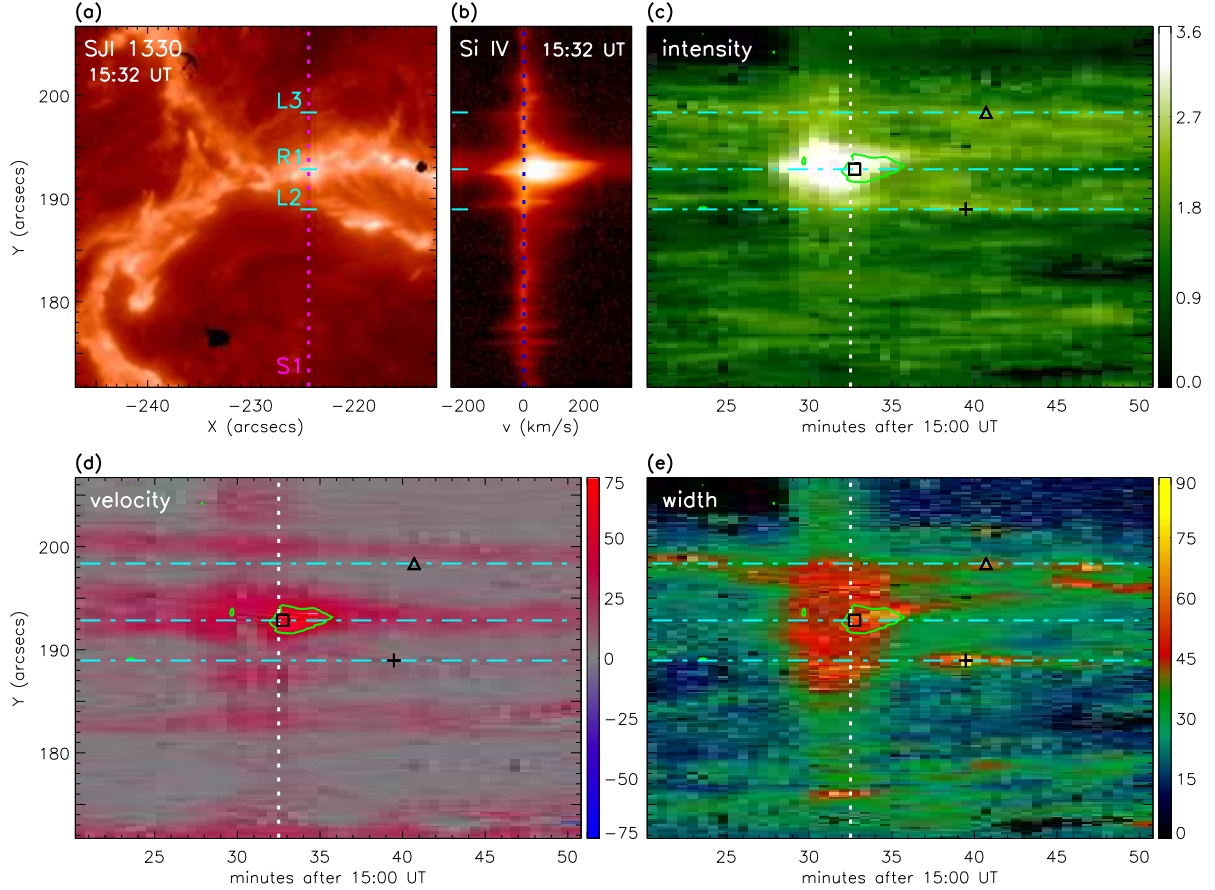}
\caption{{\small Moment analysis of the Si {\sc iv} spectra along S1. (a) SJI 1330 \AA~at the flare peak time with the magenta dotted line denoting S1. (b) Si {\sc iv} spectra along S1. The blue dotted line corresponds to the reference wavelength of the Si {\sc iv} line. (c)--(e) Temporal evolutions of total intensity (in units of log (DN s$^{-1}$)), line shift (in units of km s$^{-1}$; positive values for redshifts and negative for blueshifts), and line width (in units of km s$^{-1}$) derived from the Si {\sc iv} spectra along S1. The green contour in each panel represents the redshift with a velocity of 50 km s$^{-1}$. The vertical dotted line indicates the flare peak time. The three horizontal dash-dotted lines denote three locations L3, R1, and L2 on the slit (marked in panel (a)). For each location, we select a sample time (marked with different symbols) when the spectral lines show certain dynamic features.}}
\label{fig-siw0}
\end{figure*}

\begin{figure*}
\centering
\includegraphics[width=16cm]{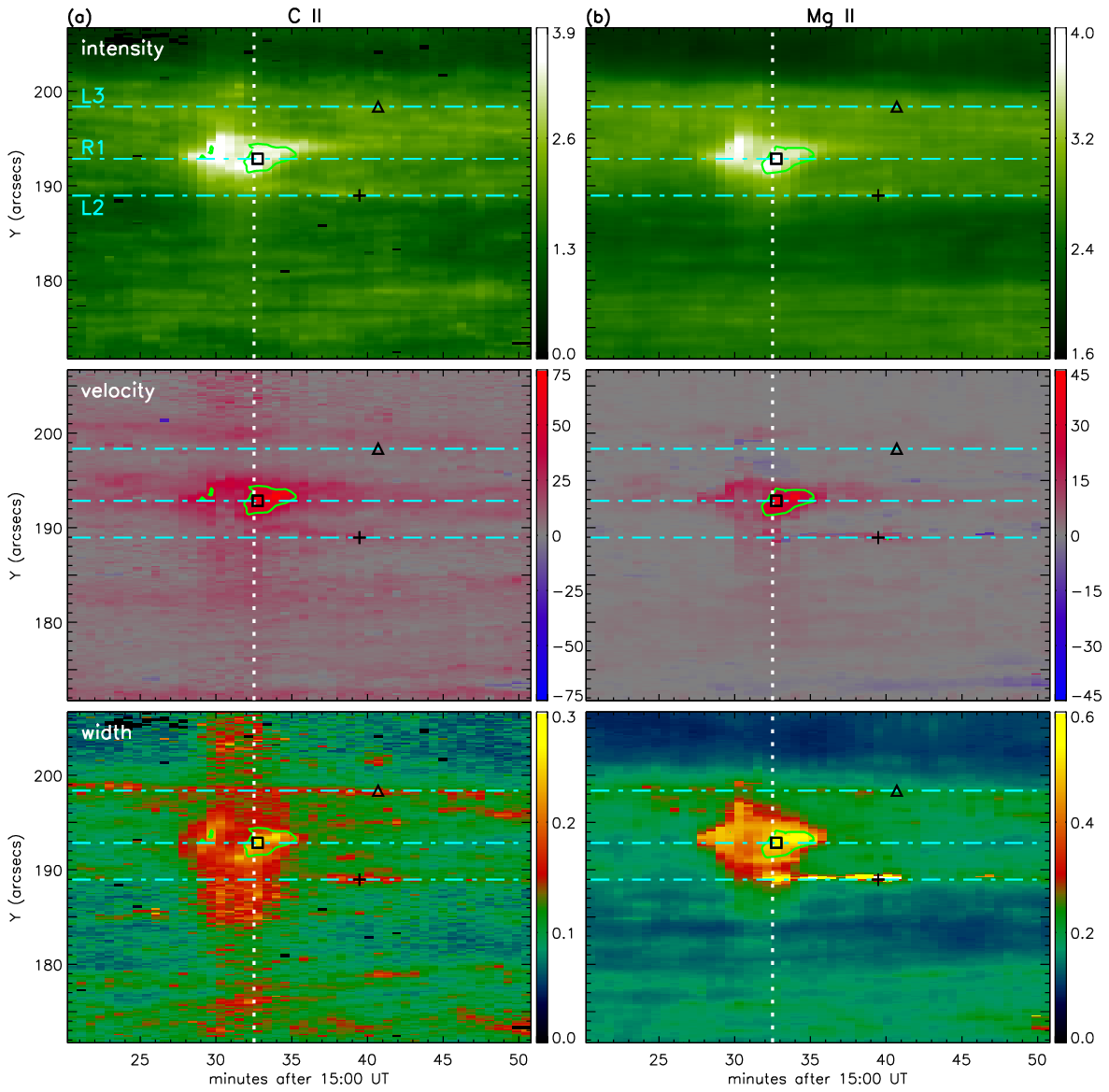}
\caption{{\small Moment analysis of the spectra of C {\sc ii} (left) and Mg {\sc ii} (right) along S1. Plotted from top to bottom are the temporal evolutions of total intensity (log (DN s$^{-1}$)), line shift (km s$^{-1}$), and line width (\AA). The green contours in panels (a) and (b) represent the redshifts with velocities of 45 and 20 km s$^{-1}$, respectively. The vertical dotted line, three horizontal dash-dotted lines, and symbols have the same meanings as in Figure \ref{fig-siw0}.}}
\label{fig-cmg0}
\end{figure*}

\begin{figure*}
\centering
\includegraphics[width=16cm]{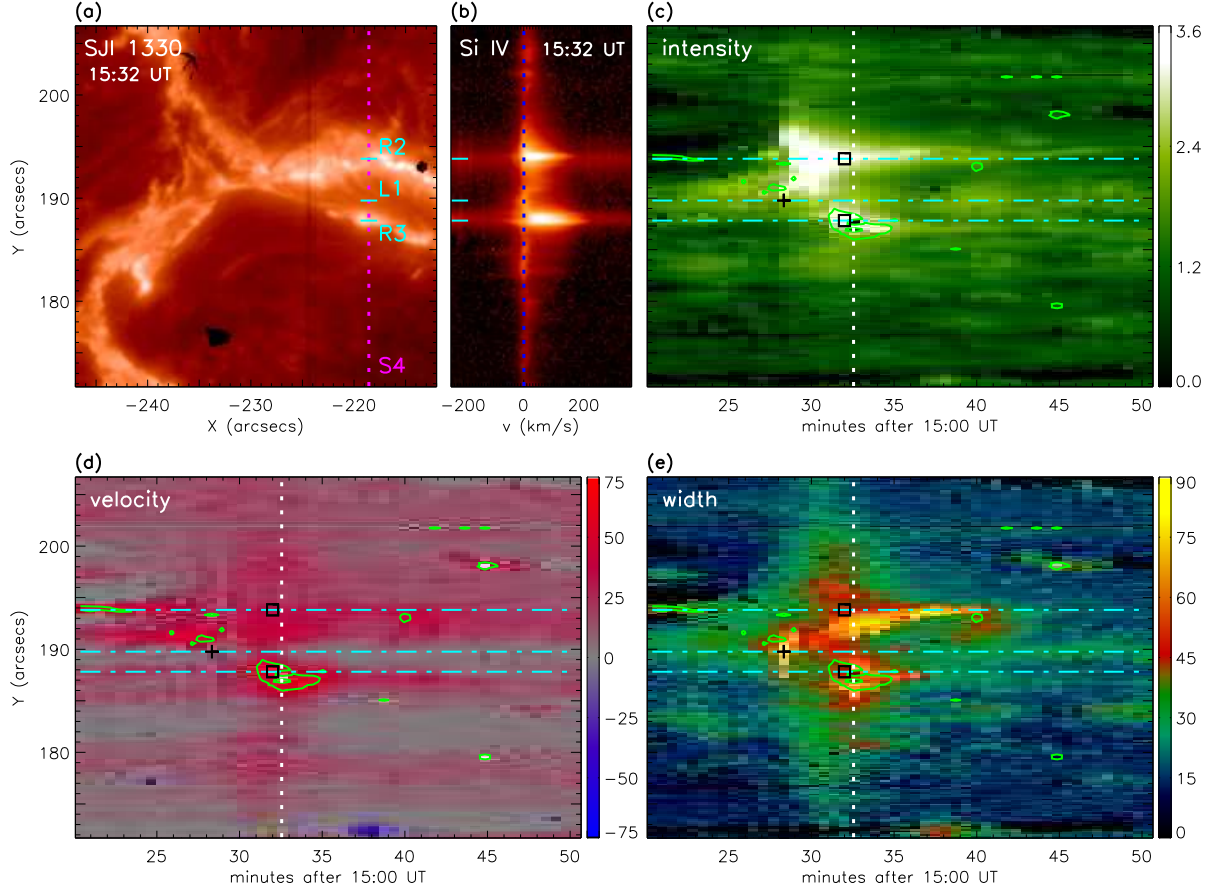}
\caption{{\small Moment analysis of the Si {\sc iv} spectra along S4. (a) SJI 1330 \AA~around the flare peak time with the magenta dotted line denoting S4. (b) Si {\sc iv} spectra along S4. The blue dotted line corresponds to the reference wavelength of the Si {\sc iv} line. (c)--(e) Temporal evolutions of total intensity (log (DN s$^{-1}$)), line shift (km s$^{-1}$), and line width (km s$^{-1}$) derived from the Si {\sc iv} spectra along S4. The green contour in each panel represents the redshift with a velocity of 50 km s$^{-1}$. The vertical dotted line indicates the flare peak time. The three horizontal dash-dotted lines denote three locations R2, L1, and R3 on the slit (marked in panel (a)). For each location, we select a sample time (marked with plus and square symbols) when the spectral lines show certain dynamic features.}}
\label{fig-siw3}
\end{figure*}

\begin{figure*}
\centering
\includegraphics[width=16cm]{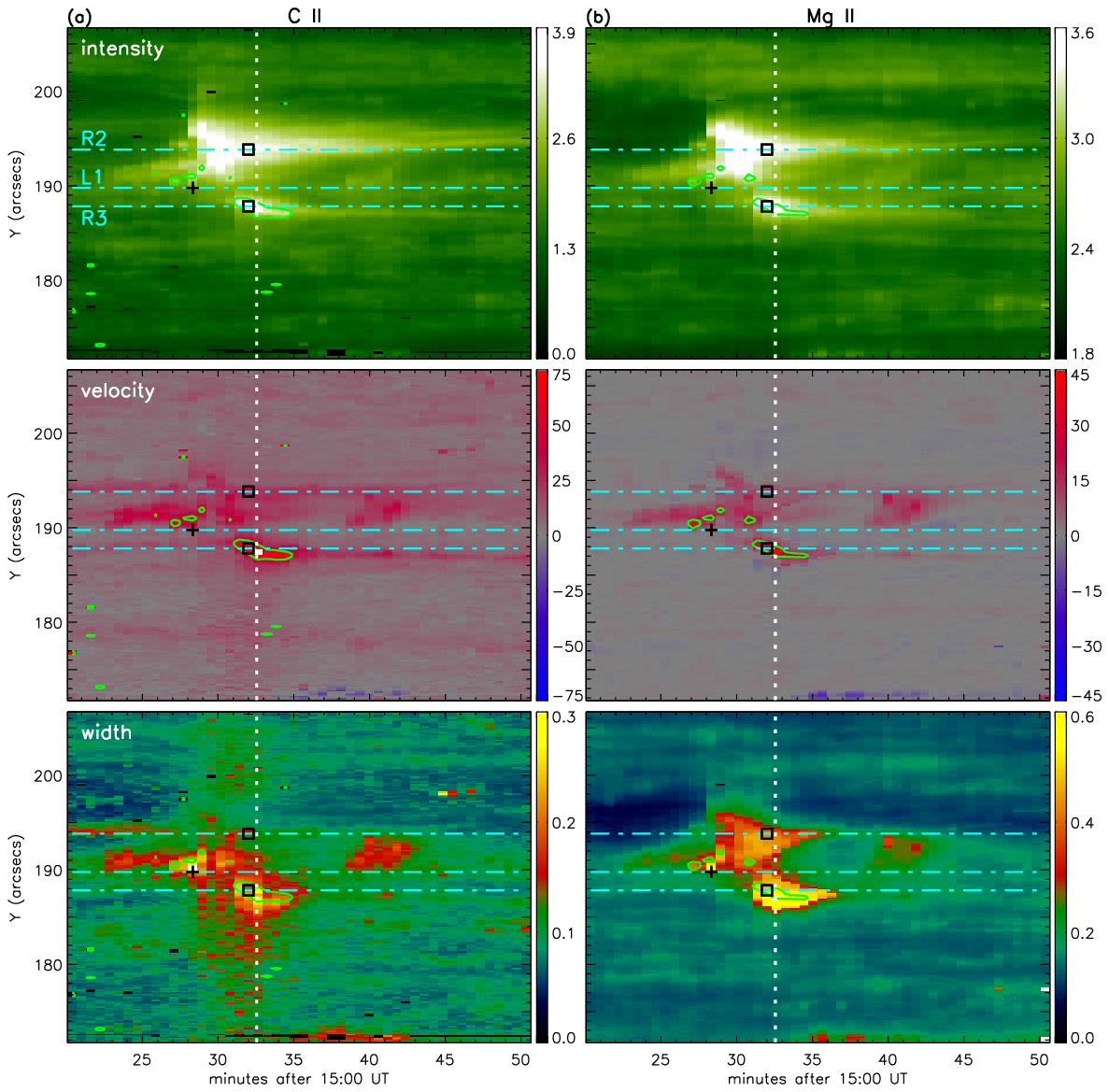}
\caption{{\small Moment analysis of the spectra of C {\sc ii} (left) and Mg {\sc ii} (right) along S4. Plotted from top to bottom are the temporal evolutions of total intensity (log (DN s$^{-1}$)), line shift (km s$^{-1}$), and line width (\AA). The green contours in panels (a) and (b) represent the redshifts with velocities of 45 and 20 km s$^{-1}$, respectively. The vertical dotted line, three horizontal dash-dotted lines, and symbols have the same meanings as in Figure \ref{fig-siw3}.}}
\label{fig-cmg3}
\end{figure*}

Based on the moment method, we generate space-time diagrams of the total intensity, line shift (or Doppler velocity), and line width of the Si {\sc iv} line for S1, as shown in Figure \ref{fig-siw0}. One can see that the north ribbon (where R1 is located) spreads up toward the north as time evolves (Figure \ref{fig-siw0}(c)). This can also be seen in the SJIs 1330 \AA~and AIA 1600 \AA~images (Animation 1). Along with the apparent motion of the ribbon, evident redshifts appear in the Si {\sc iv} line as revealed on the velocity map (see the green contour in Figure \ref{fig-siw0}(d)). These redshifts also match some broadenings of the line (see the same green contour in Figure \ref{fig-siw0}(e)). Note that there are some other brightenings, redshifts, and broadenings around R1 before the flare peak time (see $\sim$15:30 UT), which are likely related to the dynamics at the east footpoint of a flux rope (marked by the black arrow in Figure \ref{fig-lps}) that erupts early in this event. Similar features are also visible in the cooler lines. From the moment maps of the C {\sc ii} and Mg {\sc ii} lines in Figure \ref{fig-cmg0}, it can be seen that evident redshifts appear and extend as the ribbon spreads (see the green contours). The redshifts are also coincident with significant broadenings in these two lines.

A little different from S1, S4 cuts both the north and south ribbons where R2 and R3 are located, respectively. The two ribbons move apart from each other as the flare evolves, which can be clearly seen from the intensity maps for S4 in Figures \ref{fig-siw3} (Si {\sc iv}) and \ref{fig-cmg3} (C {\sc ii} and Mg {\sc ii}). Along with this, two redshift (and broadening) bands spread out. Such outward motion is more significant at the south ribbon (see the green contours in these figures). We find that the lifetime of such dynamic features, i.e., simultaneous redshifts (mostly $>$20 km s$^{-1}$) and line broadenings (above 50\% of the maximum), particularly in the Si {\sc iv} line, is about 1--8 minutes at a given site and the spatial scale is about 1--5 arcsecs along the slit.

\subsubsection{Line Profiles at R1--R3}

\begin{figure*}
\centering
\includegraphics[width=12cm]{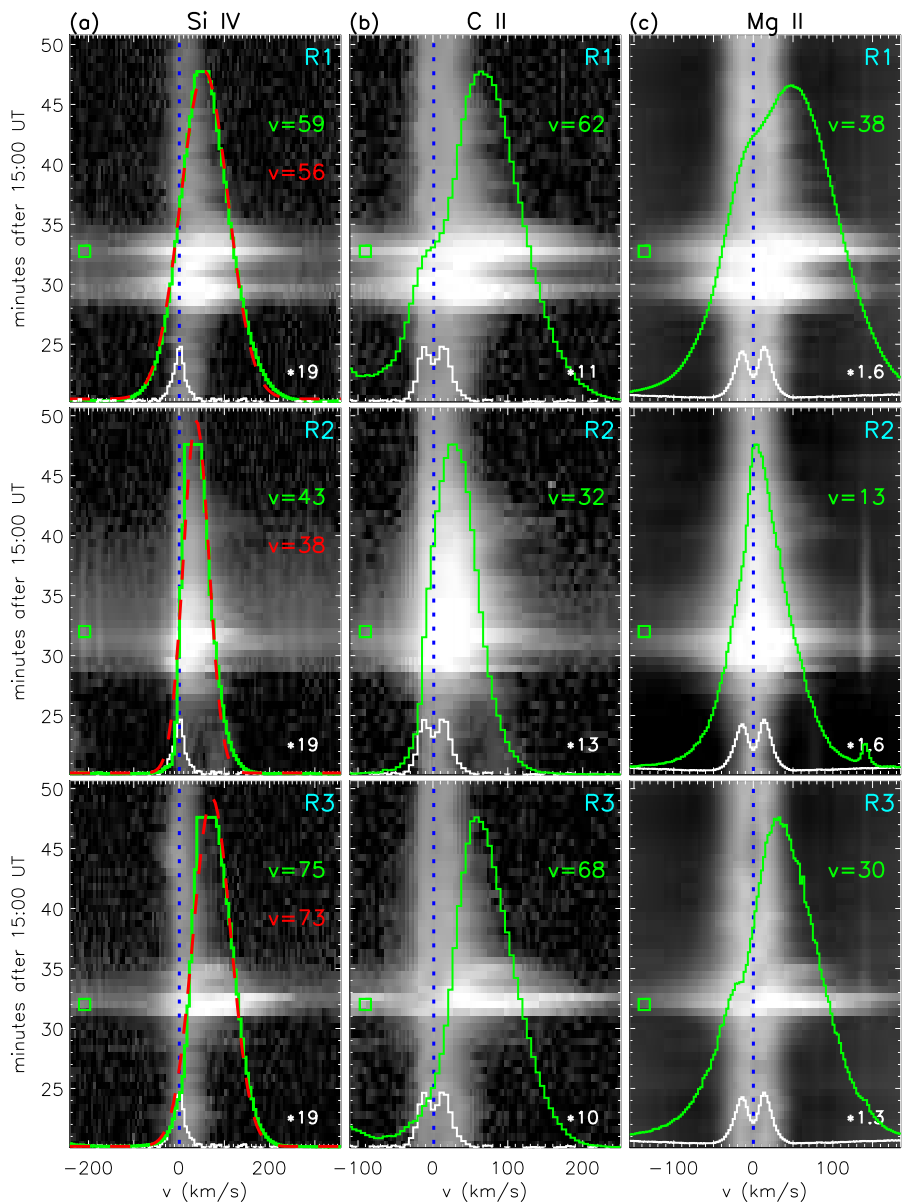}
\caption{{\small Temporal evolution of the spectra of Si {\sc iv} (left), C {\sc ii} (middle) and Mg {\sc ii} (right) at the three ribbon locations R1--R3, shown as the black-white images. Over-plotted are some featured line profiles (in green) at the times marked by square symbols. The Doppler velocities from moment analysis are given in each panel (positive values for redshifts). Note that the Si {\sc iv} line profile can be well fitted by a Gaussian function (in red), yielding a Doppler velocity as well. The vertical blue dotted line in each panel represents the reference wavelength of the corresponding spectral line. The line profile plotted in white refers to a typical profile from a relatively quiet region, which is multiplied by a factor shown in the right-hand corner of each panel.}}
\label{fig-pfr}
\end{figure*}

\begin{figure*}
\centering
\includegraphics[width=16cm]{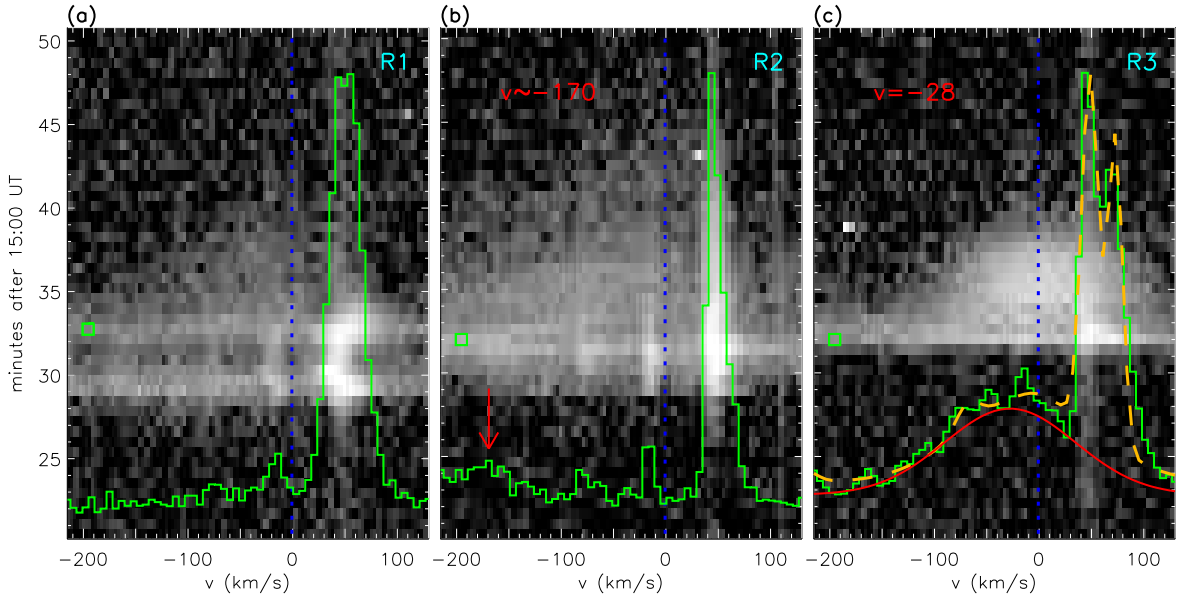}
\caption{{\small Temporal evolution of the Fe {\sc xxi} spectra at the three ribbon locations R1--R3, shown as the black-white images. Over-plotted are some featured line profiles (in green) at the times marked by square symbols. The vertical blue dotted line represents the reference wavelength of the Fe {\sc xxi} line. For the line profile at R3 showing significant Fe {\sc xxi} emission, we use a multiple Gaussian function to fit and plot in an orange dashed curve. The red solid curve represents the Fe {\sc xxi} component. The derived Doppler velocity of the Fe {\sc xxi} line is $-$28 km s$^{-1}$ (negative values for blueshifts). The red arrow in panel (b) denotes the probable Fe {\sc xxi} emission with a blueshift velocity of $\sim$170 km s$^{-1}$.}}
\label{fig-fer}
\end{figure*}

We further present temporal evolution of the spectra at three ribbon locations, i.e., R1 on S1 and R2 and R3 on S4, and show typical line profiles with prominent dynamic features at some selected times (marked by square symbols in Figures \ref{fig-siw0}--\ref{fig-fer}). The top panels of Figure \ref{fig-pfr} show the results for the Si {\sc iv}, C {\sc ii}, and Mg {\sc ii} lines at R1. It is seen that all these cool lines are brightened, redshifted, and broadened around the flare peak time (denoted by a square). The line profiles at the peak time (15:32 UT) are over-plotted with green solid curves. We can see that the Si {\sc iv} profile is Gaussian-like and redshifted as a whole, which can be well fitted by a single Gaussian function (red dashed curve). The derived velocity from the Gaussian fitting is 56 km s$^{-1}$, very similar to the value of 59 km s$^{-1}$ derived from the moment method. The C {\sc ii} and Mg {\sc ii} lines also show significant redshifts with velocities of 62 and 38 km s$^{-1}$, respectively. In particular, these two line profiles do not show a central reversal that is a common feature in the observed C {\sc ii} and Mg {\sc ii} profiles in quiet-Sun regions (see the white curves). All these spectral features at R1, including entirely redshifted Si {\sc iv} and singly peaked C {\sc ii} and Mg {\sc ii}, are also shown in the line profiles at R2 and R3, as plotted in the middle and bottom panels of Figure \ref{fig-pfr}. Note that the Si {\sc iv} line is slightly saturated at these ribbon locations.

We also check the hot Fe {\sc xxi} line at R1--R3 as shown in Figure \ref{fig-fer}. It is seen that the Fe {\sc xxi} emission is not evident at R1 at 15:32 UT, which might be hidden in the enhanced continuum background; yet this hot emission is clearly visible at R2 and particularly at R3. We then use a multiple Gaussian function to fit the line profile at R3 and obtain a blueshift velocity of 28 km s$^{-1}$ for Fe {\sc xxi}. Note that the Fe {\sc xxi} emission at R2 (marked by the red arrow) corresponds to a strong blueshift velocity of $\sim$170 km s$^{-1}$. In addition, we notice that all of the three locations show an enhanced continuum emission at 15:32 UT, which is a common feature for flare ribbons.

\subsubsection{Interpretation and Discussion}

The blueshifts in the hot Fe {\sc xxi} line and in particular, the evident redshifts in the cool Si {\sc iv}, C {\sc ii}, and Mg {\sc ii} lines, along with the ribbon spreading and line broadenings in the impulsive phase of the flare are well consistent with the scenario of chromospheric evaporation/condensation (e.g., \citealt{tian15,ying15}) caused by an energy deposition at the flare ribbons. The existence of both blueshifts (indicative of upflows) and redshifts (downflows) revealed by different spectral lines also suggests that an explosive evaporation occurs in this flare.

It is interesting that in this X-shaped flare the Si {\sc iv} line profiles at the ribbons are redshifted as a whole and thus can be well fitted by a single Gaussian function. This result is somewhat different from previous studies. \cite{ying15} analyzed an X1.0 flare that was also observed by \IRIS~on 2014 March 29 and found that the Si {\sc iv} profiles at four ribbon pixels (see the top right panels of Figures 3--6 in their paper) exhibited a redshifted component plus a rest component, which were better fitted by a double Gaussian function. \cite{tian15} also reported that the Si {\sc iv} line is often not entirely redshifted but just shows an evident red-wing enhancement at the ribbons. It is worth mentioning that some fully redshifted Si {\sc iv} profiles were reported by \cite{warr16} and also reproduced in numerical simulations by \cite{reep16}. However, those Si {\sc iv} profiles generally exhibit multiple components and might be better fitted by a double or multiple Gaussian function. The result of the entirely redshifted Si {\sc iv} line reported here implies that this line is formed within a layer (around the transition region) that is moving downward as most probably the chromospheric condensation, and absence of a rest component suggests that \IRIS~may spatially resolve the condensation region in this particular flare. Another possibility is that almost all the plasma at the transition region temperature is pushed downward, presumably by the overpressure of a local energy deposition, which might imply that considerable energy is deposited around the narrow transition region. We notice that in this X-shaped flare, the ribbons near the X-point are not correlated with any evident non-thermal hard X-ray emission; while in the X1.0 flare in \cite{ying15} and also the events in \cite{tian15} and in \cite{warr16}, hard X-ray sources are co-spatial with the ribbons. Thus, we speculate that the released energy near the X-point in this flare might be more thermal and deposited primarily in a relatively higher and narrower layer (say, the transition region) as compared with the non-thermal case (usually in the chromosphere). The shape of the Si {\sc iv} line profiles, i.e., wholly shifted or not, and their potential relation to the energy deposition will be studied in detail in more flare events as well as by numerical simulations in the future.

In the X-shaped flare, we find that the time and spatial scales of the ribbon dynamics are about 1--8 minutes  and 1--5 arcsecs, respectively. The spatial scale is similar to that of the X1.0 flare (2--3 arcsecs) reported by \cite{ying15} but the time scale is a little longer than the one (1--2 minutes) reported in that flare. As pointed out by \cite{ying15}, these scales are determined by several factors, such as the spread speed of flare ribbon, the duration of energy deposition, the hydrodynamic time scale, and the temporal and spatial resolutions of observations. We find that in this X-shaped flare, the spread speed of the flare ribbons, especially for the north and south ribbons cut by S4, is $\sim$15 km s$^{-1}$ around the flare peak time and decreases to several km s$^{-1}$ in the late decay phase. These speeds are smaller than the apparent speed of ribbon front in the X1.0 flare, which is $\sim$20 km s$^{-1}$. The smaller speeds here may lead to a narrower band, say $\sim$1 arcsec, with dynamic features in the late decay phase. The relatively longer time scale may be explained as follows. Firstly, the slight drift of the \IRIS~slit during the observations may have caused the time scale of dynamic features to be a little bit longer than that of actuality. Secondly, the energy is released and deposited on many small-scale strands within a single \IRIS~pixel. This multi-thread scenario was modeled by \cite{reep16} and the authors reproduced a long duration redshift of the Si {\sc iv} line. Finally, we do not exclude the possibility that a longer duration of energy deposition occurs in this X-shaped flare as compared with the X1.0 flare in \cite{ying15}. We notice that the time scale of dynamics in the Si {\sc iv} line appears to be longer than the time scale of energy deposition in most flare models. Nevertheless, the lasting redshifts of up to 8 minutes at some ribbon locations are most likely contributed by chromospheric condensation flows rather than cooling downflows as discussed in \cite{bros03} and \cite{tian15} for the following reasons. (1) These redshifts are co-spatial with the ribbon brightenings (i.e., signatures of magnetic reconnection) as represented by the enhancements in SJI 1330 \AA~and AIA 1600 \AA, which continue to show up until 15:42 UT (see Figure \ref{fig-fpt}(a)). (2) The redshifts are accompanied by hot Fe {\sc xxi} emissions that exhibit blueshifts (indicative of chromospheric evaporation; as seen in Figure \ref{fig-fer}(b)). Note that some redshifts and also brightenings are still visible in the Si {\sc iv} line in the late decay phase (for example, after 15:42 UT), which might be unrelated to chromospheric condensation and caused by cooling downflows.

\subsection{Spectral Features Outside the Flare Ribbons}
\label{sec-rec}

As described above, evident redshifts along with significant broadenings in the cool Si {\sc iv}, C {\sc ii} and Mg {\sc ii} lines show up at the flare ribbons around the flare peak time, indicative of chromospheric condensation. Meanwhile, significant line broadenings but no evident intensity enhancements or line shifts appear in some other places outside the flare ribbons such as at locations L1--L3. From Figures \ref{fig-siw0}--\ref{fig-cmg3}, it is seen that L1 (L2) shows clear broadenings before (after) the flare peak time (indicated by plus symbols) and L3 displays intermittent broadenings throughout the flare observations. We examine these broadened line profiles (Figure \ref{fig-pfl}) and find that their line cores are almost unshifted and that both of their line wings are markedly enhanced extending to hundreds of km s$^{-1}$ particularly in the Si {\sc iv} line. Note that at L1--L3, it is hardly seen any hot Fe {\sc xxi} emission (see Figure \ref{fig-fel}) when the cool lines show broadened wings.

\subsubsection{Broadened Line Wings}

\begin{figure*}
\centering
\includegraphics[width=12cm]{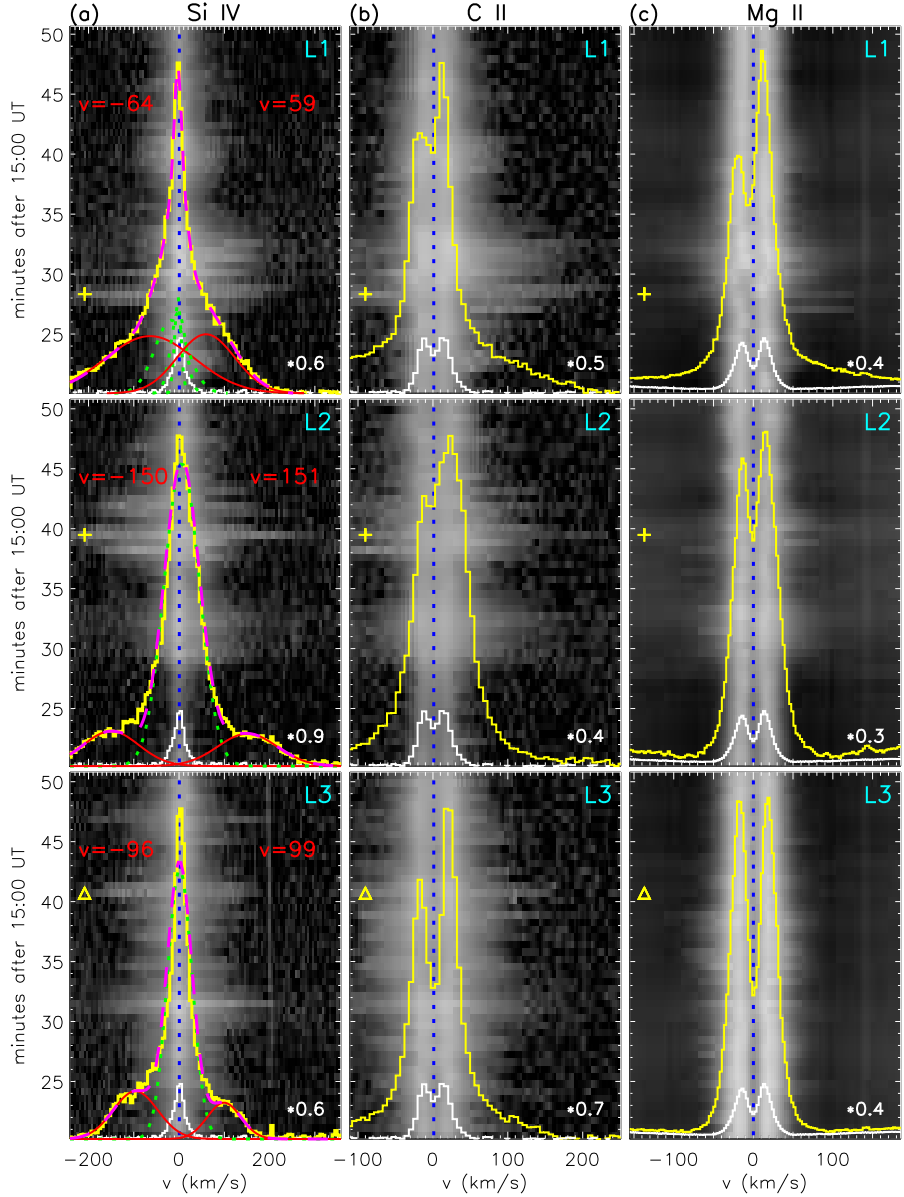}
\caption{{\small Temporal evolution of the spectra of Si {\sc iv} (left), C {\sc ii} (middle) and Mg {\sc ii} (right) at the three locations L1--L3 outside the flare ribbons, shown as the black-white images. Over-plotted are some featured line profiles (in yellow) at the times marked by plus and triangular symbols. For the Si {\sc iv} line profiles, we apply a multiple Gaussian fitting (5-Gaussian for L1 and 3-Gaussian for L2 and L3). The total fitting is plotted by a magenta dashed curve and the components are plotted by red solid curves (blueshifted or redshifted components) and green dotted curves (relatively stationary components). The velocities for the blueshifted and redshifted components are given in the Si {\sc iv} panels. The vertical blue dotted line in each panel represents the reference wavelength of the corresponding spectral line. The line profile plotted in white refers to a typical profile from a relatively quiet region, which is multiplied by a factor shown in the right-hand corner of each panel.}}
\label{fig-pfl}
\end{figure*}

\begin{figure*}
\centering
\includegraphics[width=16cm]{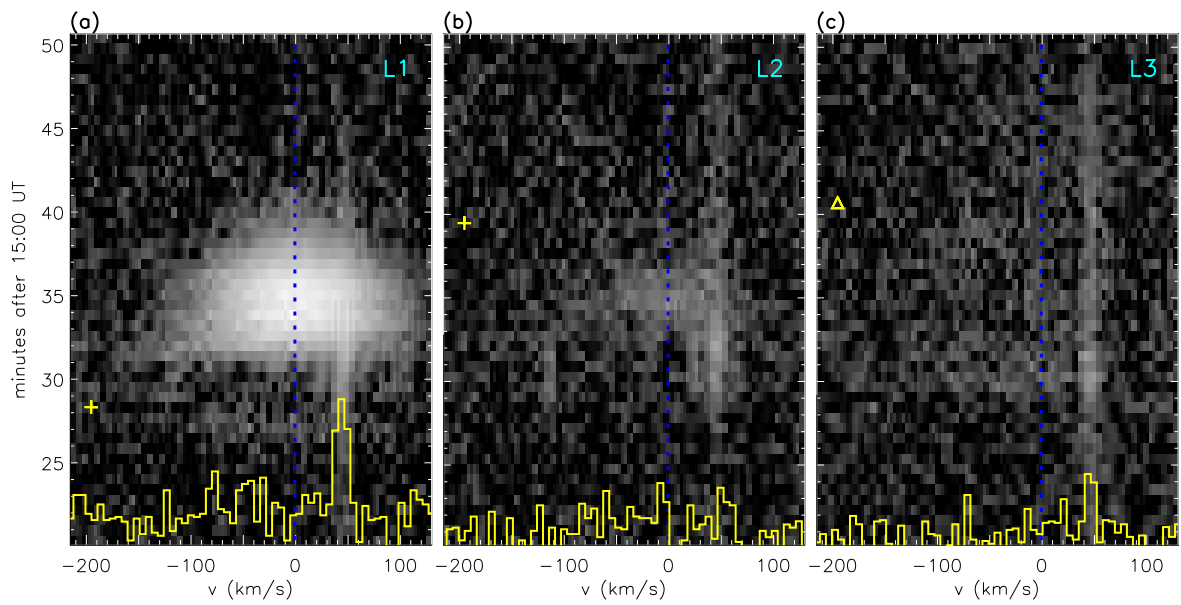}
\caption{{\small Temporal evolution of the Fe {\sc xxi} spectra at the three locations L1--L3 outside the flare ribbons, shown as the black-white images. Over-plotted are some featured line profiles (in yellow) at the times marked by plus and triangular symbols. The vertical blue dotted line represents the reference wavelength of the Fe {\sc xxi} line. It can be seen that no enhanced Fe {\sc xxi} emission shows up in these line profiles.}}
\label{fig-fel}
\end{figure*}

From Figures \ref{fig-siw3} (for Si {\sc iv}) and \ref{fig-cmg3} (for C {\sc ii} and Mg {\sc ii}), it is seen that all the cool lines exhibit evident broadenings at L1 several minutes before the flare peak time. The temporal evolutions of the spectra at L1 are plotted in the top row of Figure \ref{fig-pfl} (black-white images). Some featured line profiles, for example at 15:28 UT (indicated by a plus symbol), are over-plotted in the figure (yellow curves). One can see that the Si {\sc iv} profile deviates from a single Gaussian shape with the line core at rest. Both the blue and red line wings are significantly enhanced and extend to 200 km s$^{-1}$. Here we use a multiple Gaussian function to fit the line profile and obtain a blueshifted component with a velocity of 64 km s$^{-1}$ and a redshifted component with a velocity of 59 km s$^{-1}$. The C {\sc ii} and Mg {\sc ii} lines also show broadened wings with an unshifted and centrally reversed core. In particular, these two profiles exhibit an extended and more intense red wing (i.e., red asymmetry).  

Broadened line wings are found at L2 as well, but a few minutes after the flare peak time. The featured line profiles, for example at 15:39 UT (indicated by a plus symbol), are given in the middle row of Figure \ref{fig-pfl}. It is seen that the Si {\sc iv} line is unshifted in the core and especially shows some bumps at the far wings, somewhat similar to the featured Si {\sc iv} profile at L1 (15:28 UT). Here we also use a multiple Gaussian function to fit the Si {\sc iv} profile and derive a blueshifted component with a velocity of 150 km s$^{-1}$ and a redshifted component with a velocity of 151 km s$^{-1}$. Moreover, the Mg {\sc ii} and especially C {\sc ii} lines at L2 exhibit a red asymmetry, which is similar to L1 as well.

Line broadenings in Si {\sc iv}, C {\sc ii} and Mg {\sc ii} are also seen at L3, however, there exist some differences among L1--L3. The line broadenings at L3 start before the flare onset and persist for a long time (about 30 minutes; see Figures \ref{fig-siw0} and \ref{fig-cmg0}), while the ones at L1 and L2 only appear during the flare and last for a relatively short time (only several minutes). We notice that the broadenings at L3 appear intermittently and can be seen from the temporal evolution of the Si {\sc iv}, C {\sc ii}, and Mg {\sc ii} spectra (see the bottom row of Figure \ref{fig-pfl}). For each spectral line, we plot a featured profile from 15:40 UT (marked by a triangular symbol). One can see that the Si {\sc iv} line shows bumps at both wings with velocities of about $\pm$100 km s$^{-1}$ from a multiple Gaussian fitting, and that the C {\sc ii} line displays an obvious red asymmetry, which actually looks quite similar to the ones at L1 (15:28 UT) as well as at L2 (15:39 UT). 

\subsubsection{Interpretation and Discussion}
\label{sec-rec-dis}

The dynamic features shown at L1--L3 are distinct from the spectral features at the flare ribbons. The broadened line wings, especially the bumps at 100--150 km s$^{-1}$, in the Si {\sc iv} line profiles cannot be caused by micro-turbulence but are very likely a result of bulk plasma flows, similar to the high-speed jets in explosive events \citep{inne97,inne15}. We propose that the Si {\sc iv} profiles with bumps at both wings indicate the existence of bidirectional (i.e., upward and downward) flows that are located closely within the formation layer of the line. In addition, the C {\sc ii} and Mg {\sc ii} profiles with a red asymmetry may imply downward flows in the formation layer of the two lines.


Based on the flare morphology, L1 and L2 are supposed to be at the location of the separator, where magnetic reconnection occurs in the X-shaped flare. The strong bidirectional flows detected at L1 before the flare peak time (15:28 UT) and at L2 after the peak time (15:39 UT) are thus likely the upward and downward reconnection outflows around the separator. From the AIA 131 \AA~images ($\sim$10 MK) at 15:28 UT as plotted in Figure \ref{fig-lps}, one can see that L1 corresponds to the hot flux rope along the line of sight. We speculate that magnetic reconnection probably occurs at a relatively low site under the flux rope where the reconnection outflows (both upward and downward) are mainly captured in the Si {\sc iv} line. Moreover, we find that the north and south ribbons of the flare seem to separate from each other starting from $\sim$15:28 UT around L1 (see Animation 1). This supports the scenario that L1 could be the reconnection site at that time. We conjecture that the strong bidirectional flows detected at L2 at 15:39 UT are also possible reconnection outflows produced in the decay phase of the flare. From Figure \ref{fig-fpt}(a), one can see that the footpoint brightenings continue to appear at the outer edge of the X-shaped ribbon until 15:42 UT, demonstrating that the reconnection is proceeding into the decay phase.

The reconstructed magnetic topology (as plotted in the left panels of Figure \ref{fig-lps}) also provides some insight into the separator reconnection at L1 and L2. It is seen that the slit at steps S1 and S4 crossed the location of the separator (denoted by the red curve) calculated from the model. Some small spatial offsets may come from the uncertainty in the connectivity of the magnetic topology and/or be caused by the projection effect. Therefore, it is conceivable that bidirectional reconnection outflows can be detected at L1 and L2.

The bidirectional reconnection outflows at L1 and L2 are mainly captured in the Si {\sc iv} line, indicating that the separator reconnection (at least part of it) most likely occurs in the transition region. The transition-region reconnection could also produce some dynamic responses in the lower atmospheric layers, such as downward outflows visible in chromospheric lines. The Mg {\sc ii} and especially the C {\sc ii} line profiles at L1 (15:28 UT) and L2 (15:39 UT) display a red asymmetry, implying that the downward reconnection outflows (or reconnection downflows) are observed in the chromosphere. 

It is worth noting that bidirectional reconnection outflows revealed in Si {\sc iv} are usually reported in small-scale explosive events but very rarely in large-scale flare events in previous studies. This is because flare reconnection generally takes place in the corona and the reconnection outflows are primarily detected in hot coronal lines including Fe {\sc xxi} and Fe {\sc xix} from SUMER, Fe {\sc xxiv} and Ca {\sc xvii} from EIS, as well as Fe {\sc xxi} from \IRIS. In this X-shaped flare, however, we detect convincing reconnection signatures of bidirectional outflows in the Si {\sc iv} line at a transition-region temperature. This is owing to the special geometry of the separator reconnection. The reconnection in this flare can happen at a very low layer as the chromospheric ribbons converge at the X-point. To the best of our knowledge, it is the first time that we have detected such reconnection signatures in the cool Si {\sc iv} line during an X-shaped flare.  


Finally, the intermittent bidirectional flows revealed in the Si {\sc iv} line at L3 could also correspond to repetitive upward and downward reconnection outflows. In fact, intermittent loop brightenings occurred around L3 (see the SJIs at 1330 \AA~in Animation 3) where flux cancellations can also be seen from the HMI magnetograms (indicated by the arrows in Animation 3). These intermittent brightenings look quite similar to the explosive events that occur repetitively due to small-scale reconnections in the transition region. Such sporadic small-scale reconnections occurring close to the X-point start before the flare onset and persist throughout the flare. We conjecture that these reconnections might play a role in triggering the flare separator reconnection by destabilizing the magnetic structure and enabling the coronal loops to flow more easily into the X-point. 

\section{Summary}

In this paper, we have presented the spatio-temporal variation of the UV spectra including the Si {\sc iv}, C {\sc ii}, and Mg {\sc ii} lines for an atypical X-shaped flare observed with \IRIS. Distinct spectral features are found at the flare ribbons and outside the flare ribbons, which could be explained by chromospheric evaporation/condensation and separator magnetic reconnection, respectively.

At the flare ribbons, evident redshifts (up to 80 km s$^{-1}$) along with line broadenings are present in the cool Si {\sc iv}, C {\sc ii}, and Mg {\sc ii} lines in the impulsive phase of the flare. Meanwhile, blueshifts are observed in the hot Fe {\sc xxi} line. These blueshifts/redshifts are well consistent with the scenario of chromospheric evaporation/condensation, and suggest an explosive evaporation occurring in the flare. We find that the dynamic features spread out in the same manner as the ribbon separation with a time scale of 1--8 minutes and a spatial scale of 1--5 arcsecs, respectively. We also find that the Mg {\sc ii} and C {\sc ii} lines are singly peaked without showing a central reversal in the line core, which is consistent with earlier studies. The interesting result is that the Si {\sc iv} line is entirely redshifted with no rest component. This is different from some previous studies and will be investigated in a future work.

More importantly, at some locations outside the flare ribbons, all the cool lines exhibit significant line broadenings along with low intensity and little line shift. In particular, the Si {\sc iv} line presents broadened blue and red wings that extend to 200 km s$^{-1}$. Such broadened wings likely indicate strong bidirectional flows, which can be interpreted as upward and downward outflows produced by the separator reconnection based on the observed SJIs and AIA images as well as reconstructed magnetic topology. This kind of spectroscopic signatures of separator reconnection are rarely reported in previous flare studies. Moreover, some intermittent bidirectional outflows are detected before and during the flare and could play a role in triggering the separator reconnection.


\acknowledgments
{\em IRIS} is a NASA small explorer mission developed and operated by LMSAL with mission operations executed at NASA Ames Research center and major contributions to downlink communications funded by the Norwegian Space Center (NSC, Norway) through an ESA PRODEX contract. {\em SDO} is a mission of NASA's Living With a Star Program. The authors thank Dana Longcope for valuable discussions and thank the referee for constructive comments to improve the manuscript. Y.L., M.D.D., and W.Q.G. are supported by NSFC under grants 11373023, 11403011, 11733003, 11233008, and 11427803, by NKBRSF under grant 2014CB744203, and by {\em ASO-S} grant U1731241. Y.L. is also supported by CAS Pioneer Hundred Talents Program, Key Laboratory of Solar Activity of National Astronomical Observatories of the Chinese Academy of Sciences (KLSA201712), and by ISSI and ISSI-BJ from the team ``Diagnosing Heating Mechanisms in Solar Flares through Spectroscopic Observations of Flare Ribbons" led by Hui Tian. The work at MSU is supported by NSF grant 1460059. Part of the work was conducted during the NSF REU Program at MSU.

\bibliographystyle{apj}

\end{document}